# Coordination protocol for inter-operator spectrum sharing in co-primary 5G small cell networks

B. Singh, S. Hailu, K. Koufos, A. A. Dowhuszko, O. Tirkkonen and R. Jäntti

Department of Communications and Networking, Aalto University, Espoo 02150, Finland

R. Berry,

Department of Electrical Engineering and Computer Science, Northwestern University, Evanston, IL 60208, USA

## Abstract

We consider spectrum sharing between a limited set of operators having similar rights for accessing spectrum. A coordination protocol acting on the level of the Radio Access Network (RAN) is designed. The protocol is non-cooperative, but assumes an agreement to a set of negotiation rules. The signaling overhead is low, and knowledge of competitor's channel state information is not assumed. No monetary transactions are involved; instead, spectrum sharing is based on a RAN-internal virtual currency. The protocol is applicable in a scenario of mutual renting and when the operators form a spectrum pool. The protocol is reactive to variations in interference and load of the operators, and shows gains in a simulated small cell scenario compared to not using any coordination protocol.

## I. Introduction

In state-of-the-art mobile communication, dedicated and exclusive spectrum access coupled with unlicensed local area solutions is the mainstream approach that national regulatory authorities use to allocate new spectrum. In exclusive access, only one operator has the right to use a dedicated licensed frequency band according to specific rules. Though exclusive spectrum access will certainly be needed in 5G mobile systems to guarantee Quality-of-Service (QoS) in wide area Radio Access Networks (RANs), other regulatory options may be needed in addition. Especially at carrier frequencies above 6 GHz, and in small cell networks, exclusive access may result in low spectrum utilization efficiency. Unlicensed access, on the contrary, offers unpredictable QoS. Enabling the high capacity and flexible usage envisioned for 5G systems thus calls for more flexible regulatory regimes where, e.g., operators may operate across various frequency bands with different authorization modes [1].

*Co-Primary Shared Access* is a complementary new alternative for spectrum sharing, where multiple operators jointly use a part (or the whole) of their licensed spectrum [1,2]. The most relevant Co-Primary Shared Access scenarios are Mutual Renting (MR) and Limited Spectrum Pool (LSP). In MR, operators have individual licenses to access exclusive frequency bands, and are mutually allowed to *rent* parts of their licensed resources to their peers upon request. In LSP, a group license is given to an operator for using a common pool of spectral resources, which is shared with a limited set of operators that have equal access rights. As the set of peer operators and the principles of spectrum usage are known beforehand, investment decisions under Co-Primary Shared Access have lower risk because the long-term share of resources can have a predictable minimum value [2].

Joint use of licensed spectrum among operators can be realized either orthogonally in time [3], frequency [4,5] or space, or non-orthogonally [6,5]. In cooperative orthogonal time domain sharing, operators with a low load can borrow their time-slots to heavily loaded operators, helping them to reduce blocking probability and frame delay [3]. An upper bound for the sum capacity of a two operator orthogonal frequency domain sharing scenario is found in [4], where operators have full access to User Equipment (UE)-specific channel quality indicators of all shared channels to perform coordinated scheduling. In [5], orthogonal sharing based on pairwise exchange of resource blocks between two operators is considered.

In non-orthogonal spectrum sharing, operators simultaneously use a common block of spectral resources, creating inter-operator interference. In [6], a cooperative game approach is proposed, which converges quickly to a value close to the Nash bargaining solution; however, it requires full knowledge of action profiles in the neighborhood of the node. Non-orthogonal inter-operator spectrum sharing in the spatial domain has been considered in [5], where Channel State Information (CSI) is exchanged among operators to implement coordinated transmit beamforming and steer their antenna beams towards the most convenient direction. In all these studies, the operators benefit from cooperative spectrum sharing; however, they need to reveal proprietary information to their competitors [5,6] or to a central entity [3,4].

Operators are competitors in nature and, therefore, the rationale to make them cooperate would be either a legal framework or self-interest. In such settings, there is no reason to assume that operators are willing to exchange proprietary information with their competitors. Moreover, operators provide differentiated services to their customers, with objectives that can be categorically different according to their business models. Thus, the optimization of a joint utility by a central entity is not realistic. Co-primary spectrum sharing between operators is thus characterized by the operators not knowing each other's optimization target or network states.

Spectrum sharing can be realized by monetizing spectrum usage, and arranging auctions to determine the spectrum management [7]. Here we are interested in adaptive spectrum use on a short timescale, e.g., related to cell load variations and changing inter-operator interference conditions. It is a non-trivial task to design efficient auction mechanisms for a limited area for a limited time, and to couple operator auction strategies to their income model. Implementing auctions also requires the involvement of a trusted third party with substantial accounting infrastructure to collect bids, track payments, etc. This adds inertia to moving towards auction-based spectrum sharing. Accordingly, we shall consider non-monetized spectrum sharing, happening directly between RANs of the operators.

The interaction between self-interested players can be modeled by non-cooperative games, where players make decisions independently. A non-cooperative one-shot game formulation for unlicensed access was discussed in [8]. The players were not constrained by any rules, and freely selected power allocation strategies. When the inter-link interference is sufficiently low, the flat power allocation over all available channels represents the unique Nash Equilibrium (NE) for one-shot spectrum sharing games with complete opponent information. In multi-operator spectrum sharing, the utilities and strategies are player-specific, and not shared among players. There may also be a set of rules, either agreed among players or enforced by a legal entity. The strategic choices of the players are bound by these rules-no deviations from the rules are allowed. For instance, the legal framework governing the operation of different systems in license-exempt bands (e.g., WiFi and Zigbee) gives equal access rights to all radio devices when complying with certain power emission levels, spectral masks, channel reservation protocols, and activity rate in case of ISM bands. Considering long-term sharing in license-exempt bands, repeated game strategies that lead to favorable NE were also discussed in [8]. These games have no a priori rules. It is assumed that nodes exchange information and agree in advance upon their operational points in terms of a power allocation across the shared frequency channels. The agreement is then enforced under the threat of punishment. Obviously, without exchanging information about network states and optimization objectives, it is not possible to identify and punish cheating, and cheating as such becomes ill-defined. Thereby, in a co-primary spectrum sharing setting, the applicability of strategies based on concepts of dishonesty and punishment may be limited.

Unlike spectrum sharing in license-exempt bands, where the number of players sharing spectral resources is indefinite, we consider a fixed and known number of players with publicly known and persistent identity. We assume no intra-RAN information exchange, except for limited message exchange among operators to realize spectrum negotiations. This requires a new interface, which may over-the-air, or over the core network. We design a coordination protocol for inter-operator spectrum sharing incorporating both operator strategies and spectrum sharing rules. The proposed protocol does not require a priori agreement about the operational points of different operators, but it leads to operational points which are better for all operators as compared to the case without spectrum usage coordination.

## II. System model

This paper considers Co-Primary Spectrum Sharing among a limited number of co-located RANs belonging to different operators. The operators' RANs are full cellular networks with multiple Base Stations (BSs). A coordination protocol tailored for spectrum sharing in small cell networks is considered. At least in the first stage, it is not likely that macro-cells would participate in spectrum sharing. Only a part or cluster of the RAN may be involved in spectrum sharing, e.g., small cells of different operators located in the same building. Also, there are no constraints regarding the deployment scenario. Small cells of different operators may offer high data rate services in different buildings, or they may have partially or fully overlapping service areas.

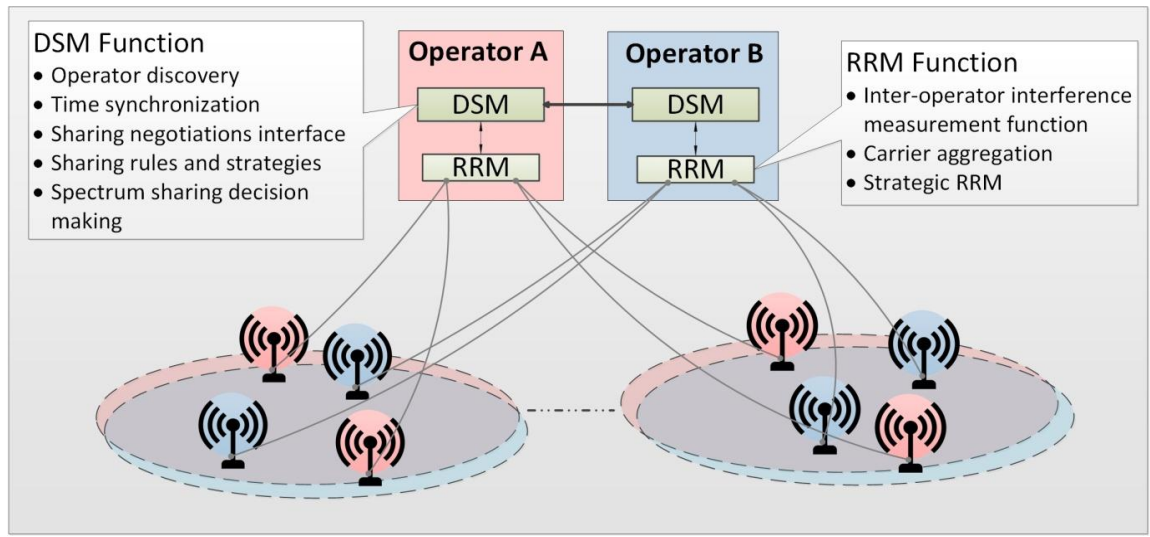

**Figure 1: Elements in system functional architecture required for inter-operator spectrum sharing on the RAN level.**

For simplicity, we concentrate on a two-operator scenario. The operators are self-interested and not willing to share operator-specific information such as load, channel usage, nor CSI. An operator is unable to reliably estimate its opponent's optimization targets and network load from RAN measurements. As a first assumption we assume that an operator ignores the state of the other operator when negotiating the use of spectrum. The decision on spectrum usage of an operator is based on its own network state, such as network load, channel conditions, UEs' locations, and interference caused by the opponent operator. Operators divide their spectrum into equal-size Component Carriers (CCs), and contribute to the spectrum sharing algorithm with an equal number of CCs. The transmit power that is used per CC is assumed constant, and spectrum sharing in the downlink is considered. This enables reliable estimation of interference caused by another operator.

We adopt a simplified version of the ETSI Reconfigurable Radio Systems (RRS) functional architecture [9] for the considered Co-Primary Spectrum Sharing, see Fig. 1. The Dynamic Spectrum Management (DSM) block is responsible for the short- and long-term management of spectrum. It decides spectrum usage and is engaged in the discovery of other operators by contacting external spectrum databases or repositories. The operators communicate only through their DSM. Coarse time synchronization is needed so that operators indicate their spectrum sharing proposals and decisions to each other almost simultaneously.

The Radio Resource Management (RRM) block performs intra-RAN interference mitigation to utilize the limited radio-frequency resources as efficiently as possible. In addition, it supports inter-operator spectrum sharing by providing inter-operator interference measurements, carrier aggregation and strategic RRM functionality for negotiating spectrum with the other operator.

## III. Coordination protocols

A coordination protocol is a mechanism to handle spectrum sharing negotiations between peer networks in inter-operator spectrum sharing. Such a coordination protocol requires: (i) A peer-to-peer connection between the operators, so that each operator can indicate its spectrum sharing proposals to the others; (ii) A set of decision rules determining the sharing outcome based on the operator proposals. These rules may, for instance, be agreed in advance between operators or enforced by an external regulator.

### A. *Spectrum sharing based on spectrum usage favors*

We assume that operators make proposals about how to share spectrum considering only their own interest, while also respecting a set of *a priori* spectrum sharing rules. A priori, the operators would be required to follow a specific MAC protocol. For simplicity, we assume that operators involved in LSP have by regulation equal access rights over the pool, and an operator should always have the right to use the full spectrum pool if desired. We shall see that with a suitable coordination protocol, a rational operator may not always use the full spectrum pool. In MR, on the other hand, each operator has the legal right to access its own spectrum, and may also give rights to other operators to use it. Thus, under MR, an operator is entitled to use its own spectrum exclusively if it is convenient from its own perspective. Hereafter, the state that an operator can take at any time instant without breaking any rules is referred to as the *fallback state*.

The coordination protocol is essentially a mechanism enabling the operator to determine the self-optimal way to use the spectrum at a particular time and announce it to the opponent. The opponent has the legal right to accept or reject the received proposal for spectrum usage. Any proposal that is accepted by the opponent necessitates a departure from the fallback state.

With concave utility functions, an accepted spectrum sharing proposal reduces the instantaneous network utility of the opponent operator and provides some instantaneous gain in the network utility of the operator that made the proposal. An accepted proposal resulting in a deviation from the fallback state of the underlying MAC protocol can be seen as a *spectrum usage favor* because the opponent is aware of its utility loss. A spectrum usage favor is exchanged only if one operator asks for it and, simultaneously, the other operator is also willing to grant it. The operators are not forced to act. Spectrum sharing based on spectrum favors entails the benefits of opponent-blind operation, non-monetized spectrum use, and a few bits of inter-operator signaling per protocol time unit.

Obviously, operators are not willing to accept everlasting performance loss in their network utilities from granting favors. The spectrum favors should be valid for a certain time period, agreed *a priori* between operators, and should be a part of the spectrum sharing rules in the coordination protocol. After the specified time period expires, the resource allocation must fall back to the state in which it was before granting the favor.

While the operator taking a favor gets an instantaneous improvement in its network utility, it is not immediately clear what is the benefit for the operator that grants a favor. As operators are self-interested entities, they will not give favors for free: some sort of fairness should be maintained. Since we do not consider inter-operator monetary transactions, the benefits from granting a favor should lie in reciprocity. It is well-known from a game theoretic perspective, e.g., in the tit-for-tat strategy, that a player will be cooperative if and only if the opponent cooperates in return. A form of reciprocity could be, for instance, that both operators give and take an equal amount of equally-valuable spectrum favors.

We discuss two coordination protocols that can be distinguished based on the reciprocity time-horizon. First, we consider the case of impatient operators that care only about instantaneous benefits. Then we focus on patient operators that are interested in long-term benefits. In both cases, we assume non-cooperative stage games played in sequence, with fixed action spaces and rules. The games are stochastic because the rewards at each stage depend on parameters governed by probability distributions, e.g., the network load, UE deployment and channel fading states. Rewards are computed with respect to the fallback state of the underlying spectrum sharing scenario.

### B. *Instantaneous reciprocity*

When the operators are impatient, we can view each stage as a separate one-shot game in which reciprocity must be satisfied. Thus, each operator gives and takes an equal number of favors at each stage game. Since the operators are selfish, a favor is exchanged only if both operators experience a positive reward at that stage.

Given the underlying spectrum sharing scenario, we next discuss strategies for the operators. The strategy is essentially the type of spectrum usage favor asked to the opponent. The target is to design strategies and spectrum sharing rules that result in favorable NE solutions.

In MR, the strategy may correspond to the amount of its own licensed spectrum the operator would like to share with the opponent, whereas in LSP, the strategy may correspond to the fraction of the pool the operator would is willing not to use. Given the spectrum sharing scenario, the operators implement their strategies independently, and exchange their proposals with the other players. After that, we need a suitable rule to resolve the spectrum sharing proposals submitted by the operators. One possibility is to select the minimum of them. Following this *minimum rule*, no operator will share more spectrum than desired under MR, whereas no operator will vacate more spectrum than desired under LSP. As a result, when an operator does not benefit from departing the fallback state, no favor is exchanged.

For concave utilities, the one-shot game following the proposed minimum rule and strategies is characterized by a unique NE [10]. In [8], spectrum sharing in the power domain was considered, and it was shown that the NE in a one-shot game is trivial with equal power on all carriers. Here, the game takes place in frequency resources, and the resulting NE may depart from full sharing. The outcome approaches the result of the cooperative games of [6].

### C. *Long-term reciprocity*

Operators are expected to share spectrum for a long time. As an operator has a persistent and publicly known identity, the operators can learn from each other's behavior. Accordingly, the interaction between operators could rather be modeled as a repeated non-cooperative game. In the repeated game, we need to keep a book of the favors exchanged because past rewards do impact future decisions. For symmetric operators, it is natural to assume that under long-term reciprocity, operators should give and take the same amount of equally valuable favors over some specified time horizon (or equivalently with some discount factor). The time horizon depends on the level of patience that operators have. Here, we consider infinitely patient operators.

In a sequence of repeated interactions over a long-time horizon, self-interested operators with no information about the RANs of other operators can develop methods to ensure cooperative gains. For instance, operators may take advantage of uncorrelated load variations in their RANs. Then, an operator with a high instantaneous load may get spectrum usage favors from an opponent operator if the opponent happens to have low load. In the future, this operator will have the chance to return these favors to show its cooperative spirit and maintain reciprocity. Even though an operator experiences performance loss by granting a favor at a stage of the repeated game, when the load situation changes, it may get a performance gain that outweighs its past loss. While exchanging favors, each operator ensures that its expected gain is larger than its expected loss, thus benefiting when compared to the case where no favors are exchanged.

In different spectrum sharing scenarios, different kinds of spectrum usage favors are asked and granted by operators. In LSP, we view a single type of favor - an operator asks the opponent for permission to use exclusively some resources from the pool. In MR, there can be bilateral agreements for resource utilization. In that case, there are two types of favors: (i) An operator asks the opponent for permission to start using jointly some of the resources of the opponent; (ii) an operator asks the opponent for permission to start using exclusively some of the resources of the opponent.

Repeated games admit a large set of equilibrium points. Since the considered game is stochastic due to the time variation of network states, it is hard to analyze and find its NE. To proceed with analysis, we resort to heuristic strategies attempting to obtain long-term reciprocity. This can be done with a threshold-based approach.

We assume that an operator knows the Probability Distribution Functions (PDFs) of its utility gains and losses. A simple long-term reciprocal strategy would be that an operator asks for a favor if its immediate utility gain is higher than a threshold $\theta_g$, and grants a favor (upon asked) if its immediate utility loss is smaller than another threshold $\theta_l$. These decision thresholds depend not only on the current network state and the gain and loss statistics, but also on the sequence of previous interactions with opponent operators. The thresholds for Operator A would be coupled to the probabilities of Operator B to grant and ask favors, so that long-term reciprocity is achieved. Details for setting the decision threshold in the case of a LSP can be found in [11]. The proposed heuristic strategy performs strictly better than the strategy that does not involve the exchange of favors. Unlike the one-shot game, where an operator takes and gives favors simultaneously, we simplify the repeated game by assuming that no action is taken when both operators ask a favor at the same stage of the game.

### D. *Combined coordination protocol*

In Co-Primary small cell deployments, the interference conditions and the RAN load are expected to vary significantly. When UEs are located close to their serving BSs and, as a consequence, the inter-operator interference is small, and both operators would be eager to share spectrum. On the other hand, when UEs are exposed to high inter-operator interference, the operators would prefer to orthogonalize their resources. Furthermore, patient operators would be willing to empty spectrum resources when they have few or no UEs to serve, and request more spectrum resources when they really need them.

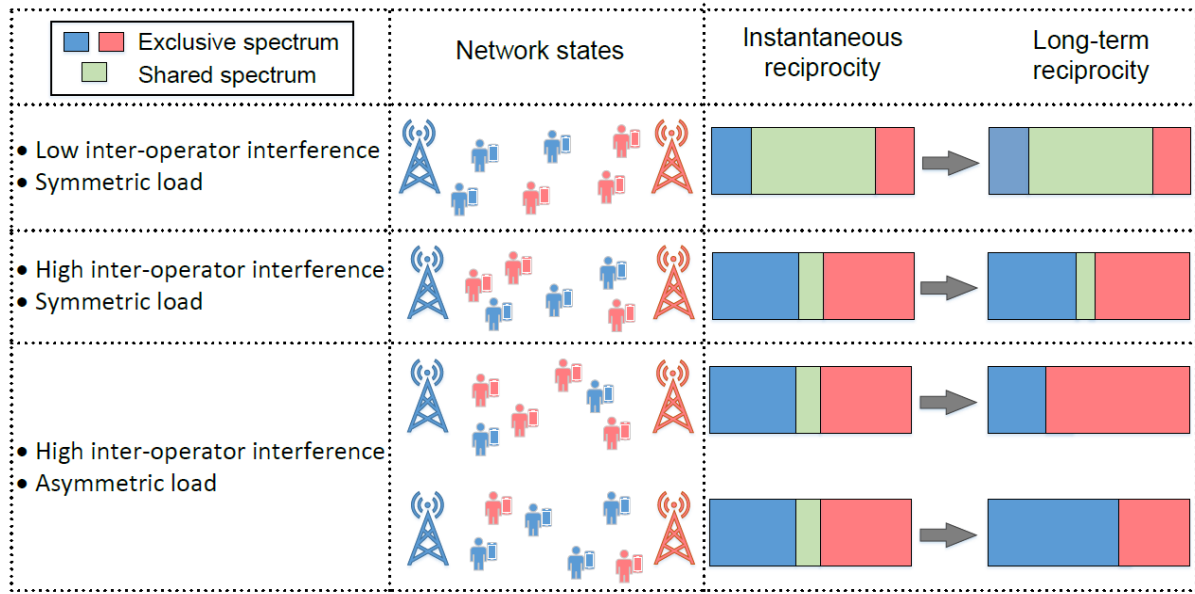

**Figure 2: Visualizing the functionality of the proposed coordination protocol based on the combined short-term and long-term reciprocity over four stage games.**

Irrespectively of the spectrum sharing scenario, the one-shot game adapts spectrum sharing to the inter-operator interference conditions. However, the outcome of the one-shot game does not depend on the RAN load variations. This is where repeated games can come into play by allocating more spectrum to the operator with higher load, provided that this operator has been cooperative in the past. We thus consider a combined protocol where in each instance of the game operators first exchange spectrum usage favors based on short-term reciprocity to adapt to the inter-operator interference situation, and then exchange favors based on long-term reciprocity, to exploit network traffic dynamics. The functionality of the proposed coordination protocol is illustrated in Fig. 2.

## IV. Numerical illustrations

We study the UE rate improvement for operators applying spectrum sharing coordination protocols in an MR scenario. We consider an indoor deployment with two LTE small cell operators in a single-story 50x50 m$^2$ building, see Fig. 3. The building has four identical rooms, and an operator's BSs are stationed in diagonal rooms. We consider downlink transmissions, a proportionally fair utility function for both operators and a full buffer traffic model. Thus the number of UEs represents the network load, which is generated from a Poisson distribution with possibly different means for different operators. The UEs are uniformly distributed in the service area of operators. Cell association is based on received signal power.

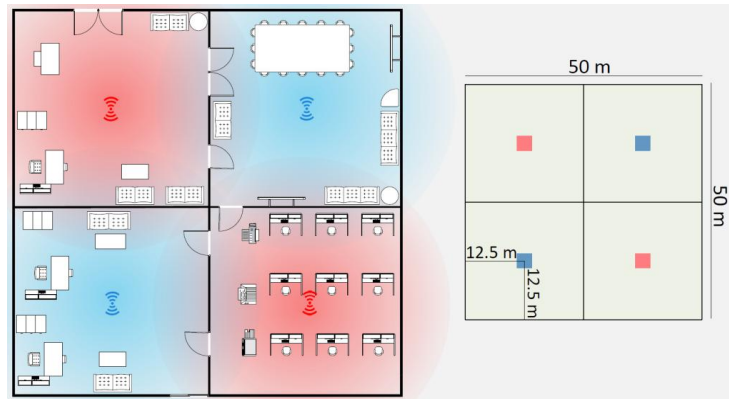

**Figure 3: Indoor inter-operator deployment scenario. Different colors represent BSs of different operators.**

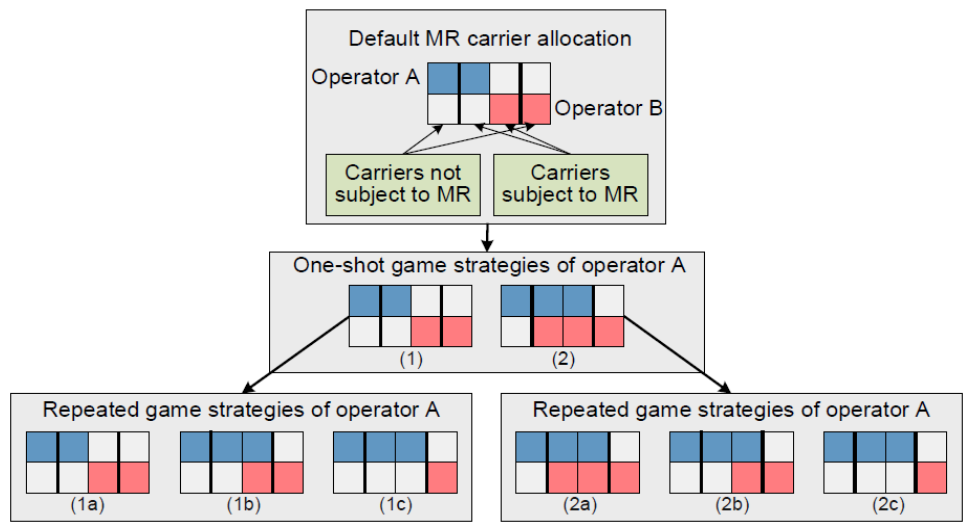

**Figure 4: Strategy profile for Operator A.**

We consider the WINNER indoor office propagation channel model [12]. A bandwidth of 80 MHz is split into four CCs, with each operator owning an exclusive license for two CCs. The available power budget per CC is 20 dBm, and the aggregate external interference plus noise level per CC is -80 dBm. Each operator contributes a CC to the spectrum sharing game and reserves the other for its own exclusive usage. Fig. 4 depicts the strategy profile for Operator A.

At each stage game, operators first play the one-shot game and select strategy (1) or (2) in Fig. 4. Based on the minimum rule, the outcome with the least carrier sharing is selected. For instance, the outcome of the one-shot game is (2) only if both operators propose to share a CC. When one or both operators are unwilling to share a CC, the outcome is (1). Next, the repeated game is executed. For example, if the outcome of the one-shot game is (2) and Operator A asks for an exclusive favor of two CCs from Operator B, which Operator B grants, the result is (2c).

We evaluate the performance of the combined coordination protocol over a finite time horizon of 4000 stage games. First we consider a scenario with equal mean network loads between the operators and low inter-operator interference. The coverage areas are non-overlapping; the UEs are served by a BS in the same room, as in Fig. 3. The mean number of UEs for each operator is 5 and the wall loss is 10 dB.

In Fig. 5, the rate distribution for the UEs of an operator is depicted. The QoS with sharing is significantly better than without. The rate improvement in 10th percentile of rate Cumulative Distribution Functions (CDF) is 47% and in 50th-percentile is 45%. The full spectrum sharing outcome (2a) is most likely. This is the ideal solution in a low interference environment. The gains of the combined protocol and the one-shot scheme are virtually the same. Only few favors are exchanged during the repeated game as the operators' loads are similar. In addition, the outcome of a fully cooperative protocol is depicted, where the operation of the two networks is jointly optimized. The full cooperation results do not differ from those of the coordination protocol.

Next, we analyze a situation with load asymmetry and high inter-operator interference. The UEs of both operators are distributed uniformly in the whole building with no internal walls. In half of the instances, the mean number of UEs is 8 for Operator A and 2 for Operator B, whereas in the other half, the loads are reversed.

In Fig. 6, we see a decline in UE rate in comparison to Fig. 5 due to intense inter-operator interference. One-shot sharing provides a small rate improvement. However, the gain is less prominent than in Fig. 5. A marginal improvement of 1% in the 10th percentile UE rate is seen, and an increment of 8% in the 50th-percentile. In the combined coordination protocol, Operator B grants more favors than Operator A when it has low load, as it can cope with fewer CCs. When it has high load, it asks and is granted for more favors. The overall performance is better than for the one-shot game. The 10th percentile UE rate is improved by 26% and the 50th percentile rate by 23% compared to no sharing. The combined coordination protocol performs close to a fully cooperative joint optimization of the networks.

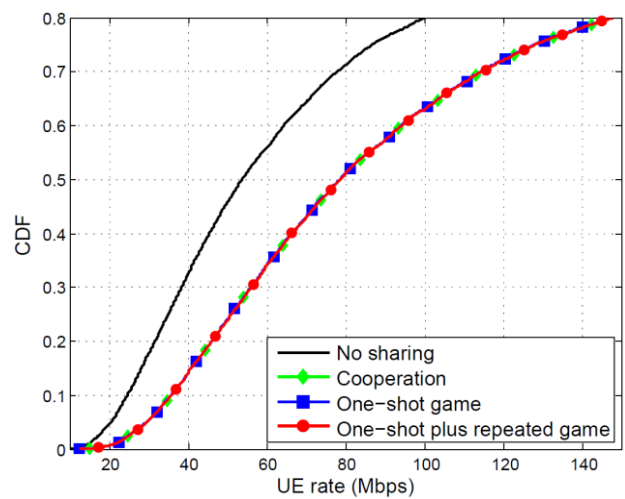

**Figure 5: Rate distribution for the UEs of an operator. Equal mean network load for the two operators, low inter-operator interference.**

In the simulation setting, UEs with low rate in no sharing are UEs in a high-load instance that are far from the serving BS and accordingly close to an opponent BS. These UEs do not benefit from the one-shot game, as they do not benefit from shared spectrum. However, the repeated game can provide more resources to these UEs, and accordingly a better rate. On the contrary, UEs that experience a high rate in no sharing are UEs close to the serving BS, and accordingly far from opponent BSs. These UEs are low-interference UEs that benefit from shared spectrum provided by the one-shot game, whereas the repeated game has little effect on their performance.

## V. Conclusions

The principle of allocating spectrum to mobile network operators based on a dedicated and exclusive license will persist, as a method to ensure coverage and Quality of Service. Nevertheless, the high capacity demands generated in hot spot areas require heterogeneous network structures. Operation of small cell networks only based on dedicated licenses may not be feasible as more spectrum is needed and new spectrum is expensive and difficult to identify. Flexible spectrum use and co-primary spectrum access is a way forward for indoor small cells. Different operators providing wireless data access in spatially separated indoor areas, or with highly directive millimeter wave technologies, may benefit from spectrum sharing due to negligible inter-operator interference. The spectrum needs for such operators would vary in space. With overlapping coverage areas, inter-operator interference may be significant, especially for centimeter-wave and lower frequencies. Load variations and changing user locations make this interference highly variable in small cell networks, so that spectrum needs of operators would vary in space and time. In order to exploit variations in spectrum needs, inter-operator spectrum sharing is a viable option. To realize it, we need a coordination protocol that has low implementation complexity, is transparent to the operator's revenue model and does not require excessive information exchange among operators. Such a protocol would also enable the emergence of new types of market players, e.g., local operators. We have designed a protocol that adapts spectrum allocation to inter-operator interference situations and network traffic dynamics which are expected to be prominent is small cells. We illustrated that in an indoor deployment scenario, two operators are both able to offer higher user rates than they could without coordination. Our results show that a rational operator, knowing that the opponent is rational and has a network with similar characteristics, has incentive to coordinate the spectrum usage. Beyond improvements in service rates, there are other issues that can influence operators' incentives to cooperate, such as the fact the service quality is a key way operators attempt to differentiate themselves from their competitors. The discussed protocol can be used in any spectrum reserved for mobile communication, for example in Licensed Shared Access [13] spectrum.

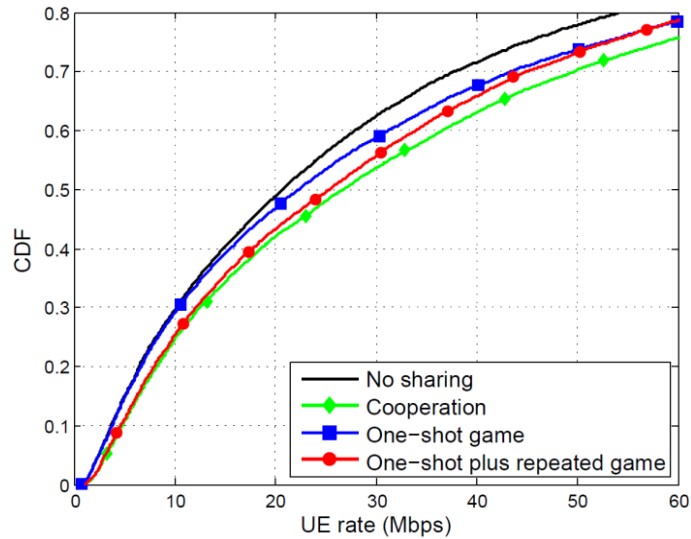

**Figure 6: Rate distribution for the UEs of an operator. Unequal mean network loads for the two operators, high inter-operator interference.**

ACKNOWLEDGMENT

This work was supported in part by the European Commission in the framework of the FP7 project ITC-317669 METIS.

BIOGRAPHIES

Bikramjit Singh (bikramjit.singh@aalto.fi) received his M.Sc. degree in Communications Engineering from Aalto University, Finland in 2014. Currently, he is pursuing a D.Sc. (Tech.) degree in Communications Engineering from Aalto University in the field of spectrum sharing in 5G heterogeneous networks.

Sofonias Hailu (sofonias.hailu@aalto.fi) received his M.Sc. degree in Communications Engineering from Aalto University, Finland in 2014. Currently, he is pursuing a D.Sc. (Tech.) degree in Communication Engineering from Aalto University in the field of 5G mobility, spectrum and radio resource management.

Konstantinos Koufos (konstantinos.koufos@aalto.fi) obtained the diploma in electrical and computer engineering from Aristotle University, Greece, the M.Sc. and the D.Sc. in radio communications from Aalto University, Finland. He is currently a Post-Doctoral researcher at the Department of Communications and Networking, Aalto University. His current research interests are in interference modeling and spectrum sharing.

Alexis Dowhuszko (alexis.dowhuszko@aalto.fi) received his Telecommunications Engineer degree from Blas Pascal University, Argentina, in 2002, and his Ph.D. degree in Engineering Sciences from the National University of Cordoba, Argentina, in 2010. He is currently a Post-Doctoral researcher at the Department of Communications and Networking, Aalto University. His research interests are in radio resource management for ultra-dense wireless networks.

Olav Tirkkonen (olav.tirkkonen@aalto.fi) received his M.Sc. and Ph.D. in theoretical physics from Helsinki University of Technology, Finland. Currently he is an Associate Professor in communication theory at the Department of Communications and Networking in Aalto University, Finland. His current research interests are in coding theory, multiantenna techniques, as well as cognitive and heterogeneous cellular systems.

Riku Jäntti (riku.jantti@aalto.fi) is an Associate Professor in Communications Engineering and the head of the department of Communications and Networking at Aalto University School of Electrical Engineering, Finland. He received his M.Sc. in Electrical Engineering in 1997 and D.Sc. in Automation and Systems Technology in 2001, both from Helsinki University of Technology. His research interests are in machine type communications, *cloud based radio access networks, spectrum and co-existence management and RF Inference*.

Randall Berry (rberry@eecs.northwestern.edu) received a M.S. and Ph.D. in Electrical Engineering and Computer Science from the Massachusetts Institute of Technology in 1996 and 2000, respectively. He is currently a Professor in the Department of Electrical Engineering and Computer Science at Northwestern University. His research interests include wireless communications and network economics.